\documentclass[sigplan,twocolumn]{acmart}
\makeatletter
\renewcommand\@subtitlefont{\Large} 
\makeatother
\renewcommand\footnotetextcopyrightpermission[1]{}
\usepackage{listings}
\usepackage{xcolor} 
\floatstyle{ruled}
\newfloat{listing}{tb}{lst}{}
\floatname{listing}{Listing}
\AtBeginDocument{%
  }

\setcopyright{acmlicensed}
\copyrightyear{2018}
\acmYear{2018}
\acmDOI{XXXXXXX.XXXXXXX}
\acmConference[Some ACM conference]{Make sure to enter the correct
  conference title from your rights confirmation email}{June 03--05,
  2018}{Woodstock, NY}
\acmISBN{978-1-4503-XXXX-X/2018/06}

\begin{document}

\title{Building a Correct-by-Design Lakehouse}
\subtitle{Data Contracts, Versioning, and Transactional Pipelines for Humans and Agents}
\author{Weiming Sheng}
\authornote{All authors contributed equally and are listed \texttt{ORDER BY AGE ASC}. JT is the corresponding author and PI on the project:  \url{mailto:jacopo.tagliabue@bauplanlabs.com}.}
\affiliation{%
  \institution{Columbia University}
  \country{USA}
}

\author{Jinlang Wang}
\authornotemark[1]
\affiliation{%
  \institution{University of Wisconsin-Madison}
  \country{USA}
}

\author{Manuel Barros}
\authornotemark[1]
\affiliation{%
  \institution{Carnagie Mellon University}
  \country{USA}
}

\author{Aldrin Montana}
\authornotemark[1]
\affiliation{%
 \institution{Bauplan Labs}
 \country{USA}
}

\author{Jacopo Tagliabue}
\authornotemark[1]
\affiliation{%
 \institution{Bauplan Labs}
 \country{USA}
}

\author{Luca Bigon}
\authornotemark[1]
\affiliation{%
  \institution{Bauplan Labs}
  \country{USA}
}

\renewcommand{\shortauthors}{Sheng et al.}

\begin{abstract}
\setlength{\emergencystretch}{2em}
Lakehouses are now the default substrate for analytics and AI, but they remain fragile under concurrent, untrusted change: schema mismatches often surface only at runtime, development and production easily diverge, and multi-table pipelines can expose partial results after failure. We present Bauplan, a code-first lakehouse that aims to eliminate a broad class of these failures by construction. Bauplan builds on a storage substrate that already provides atomic single-table snapshot evolution, and adds three pipeline-level correctness mechanisms: typed table contracts to make transformation boundaries checkable, Git-like data versioning to support reproducible collaboration and review, and transactional runs that guarantee atomic publication of an entire pipeline execution. We describe the system design, show how these abstractions fit together into a unified programming model for humans and agents, and report early results from a lightweight Alloy model that both validates key intuitions and exposes subtle counterexamples around transactional branch visibility. Our experience suggests that correctness in the lakehouse is best addressed not by patching failures after the fact, but by restricting the programming model so that many illegal states become unrepresentable.
\end{abstract}



\received{30 January 2026}

\maketitle

\section{Introduction} \label{sec:intro}

The data lakehouse is today the \textit{de facto} cloud architecture for analytics and AI workloads, combining object storage with open table formats and decoupled, multi-language compute \cite{Zaharia2021LakehouseAN,Tagliabue2023BuildingAS,mazumdar2023datalakehousedatawarehousing}. As the rise of coding agents is taking industry by storm \cite{huang_control_2025}, AI adoption on data workloads lags behind, as experts point out that the affordances in traditional OLAP systems make LLMs fundamentally unsafe \cite{tagliabue2025trustworthyaiagenticlakehouse}: as the majority of labor shifts from writing code to \textit{verifying and approving changes}, correctness in the face of untrusted actors becomes non-negotiable for data workloads.

In \textit{this} paper, we share the design of \textbf{Bauplan} and report preliminary results in formalizing its correctness. Bauplan is a novel lakehouse designed with vertically integrated APIs for programming, concurrency and transactions: Bauplan users (humans \textit{and} agents alike) are presented with familiar abstractions (e.g. typed functions, commits, branches), only slightly restricted in scope to allow correctness guarantees at the system level. In particular, we propose an (almost) \textit{correct-by-design} lakehouse by exploring the design space along three axes, inspired by software engineering best practices:

{\setlength{\emergencystretch}{2em}
\begin{enumerate}
  \item \textbf{data contracts through type checks}: as static checks prevent many software bugs, lightweight table contracts should prevent interface mismatches between pipeline nodes;
  \item \textbf{collaboration through ``Git-for-data''}: when Git-based primitives are applied to data assets, point-in-time debugging and pull requests (\textit{PRs}) enable human-in-the-loop verification;
  \item \textbf{correctness through transactional pipelines}: Bauplan automatically snapshots \textit{reads} and sandboxes \textit{writes} in a given pipeline execution, guaranteeing pipeline atomicity: downstream systems observe either all outputs of a run or \textit{none}.
\end{enumerate}
}

Taken together, these primitives make pipelines statically checkable, safe to run, easy to review and straightforward to reproduce. While contributions in each of the above dimensions can be enjoyed independently, we contend that our most original contribution is \texttt{Bauplan} \textit{itself}, as a physically distributed system governed by a handful of logically atomic APIs designed to maximize verifiability and correctness. Notwithstanding agentic coding as a core use case from our growing user base, our design sits more generally at the intersection of programming languages, data management and distributed systems: as such, we believe our lessons from the trenches to be valuable to a broad set of practitioners.

\section{Lakehouse failure modes: an industry perspective}
\label{sec:failures}

Data pipelines -- DAGs of transformations moving, cleaning, transforming raw tables into intermediate and final assets for BI, ML and AI consumption -- are the most significant use case for the lakehouse, from cloud usage, developer time and correctness perspective \cite{cdms2025eudoxia,Renen2024}. As a recurrent example, let us introduce the following building blocks of a sample pipeline:

\begin{lstlisting}[
  language=SQL,
  showstringspaces=false,
  columns=fullflexible,
  caption={Node \#1 as a SQL transformation},
  basicstyle=\ttfamily\scriptsize,
  numbers=none
]
-- parent.sql
SELECT col1, col2, SUM(col3) as _S FROM raw_table
\end{lstlisting}

\begin{lstlisting}[
  language=Python,
  showstringspaces=false,
  columns=fullflexible,
  caption={Node \#2 and \#3 as Python transformations},
  basicstyle=\ttfamily\scriptsize,
  numbers=none
]
def child(df=parent):
    # do something with the input table and return a table
    return df.do_something()

def grand_child(df=child):
    # do something with the input table and return a table
    return df.do_something()
\end{lstlisting}

The intended semantics respects standard data engineering conventions: \texttt{raw\_table} in the data lake gets transformed by a SQL query into a \texttt{parent} table, which gets fed into the downstream \texttt{child} transformation, this time expressed as a Python function. Finally, \texttt{grand\_child} concludes the DAG. By design, the DAG is only sketched with basic code constructs, so that our considerations can remain agnostic as to the specific implementation of the pipeline in existing platforms -- importantly, Section~\ref{sec:programming} shows how Bauplan's users need only to add a few annotations to such a vanilla scaffolding to obtain important correctness guarantees.

We highlight important failure modes in traditional platforms following our tripartite distinction from Section \ref{sec:intro}. Note that while these failure modes are already challenging for workloads run by trusted human experts, their severity is amplified in a world in which unsafe, untrusted agents are allowed to operate on production data \cite{tagliabue2025safeuntrustedproofcarryingai}:

\begin{enumerate}
    \item \textbf{schema failures}: a large fraction of pipeline errors is due to schema changes at the intersection of two nodes \cite{FOIDL2024111855}, as columns get dropped or replaced, types change, semantics shift. For example, if \texttt{col3} becomes a \textit{float} in \texttt{raw\_table}, the SQL node will still run, but break code in \texttt{child} that assumes an \textit{int};
    \item \textbf{collaboration failures}: the PR-based flow is a tried-and-tested method to add verification and peer-review to the development process \cite{10.1109/TSE.2016.2576451}. The analogy with data teams breaks down as pipeline code is inherently stateful, as its effects depend on the state of the data lake as well: a new query variation for \texttt{parent} may work in ``dev'' but not in production due to data skew; moreover, a daily pipeline that failed yesterday may succeed today even with \textit{no code changes}, breaking the standard software assumption of Git-based reproducibility;
    \item \textbf{correctness failures}: even if a lakehouse guarantees table-level atomicity through open formats \cite{iceberg_spec}, a pipeline is crucially multi-table and often multi-language. In the absence of a notion of ``transactional pipeline'', downstream readers may observe a mix of old and new tables with no clear notion of whether it completed. Notably, at the moment of drafting \textit{this} paper, the leading industry lakehouses (\textit{Snowflake} and \textit{Databricks}) do \textit{not} offer any ready-to-use API for pipeline transactions.
\end{enumerate}

Obviously, we could try a piecemeal approach to patch the situation. For example, we could maintain a so-called ``business data catalog'' to detail our data assets; however, business catalogs require a large effort to build and maintain \cite{9140254}. Even if built, it is not clear how that knowledge would easily percolate to coding assistants. We could then devise a syncing mechanism that periodically aligns ``dev'' and ``prod'' in an effort to reduce data skews; however, syncing processes are expensive to maintain, and won't eliminate skew within a finer temporal resolution. Finally, we could rely on an outer layer to provide pipeline rollbacks \cite{prefect}; however, leaving transactions to the application developer is notoriously the source of many correctness issues \cite{10.1145/3638553}. Moreover, it would result in a much steeper learning curve: the success of MVCC-style transactional guarantees \cite{10.5555/12518} is in no small measure due to the ``hands-off'' nature of SQL abstractions.

Unsurprisingly, researchers started to question the possibility that patching traditional platforms could deliver a verifiable and correct lakehouse design in a timely fashion \cite{liu_agentfirst_2025}. The common root cause is semantic: existing lakehouses provide table-level atomicity, but expose no pipeline-level programming model that makes interfaces checkable, collaboration reproducible, and multi-table effects atomic. We therefore design Bauplan around three minimal, composable primitives to eliminate these failure modes by construction: typed contracts, Git-like data versioning, and transactional runs.

\section{Design principles}
\label{sec:design}

\begin{figure}
\centerline{\includegraphics[width=0.45\textwidth]{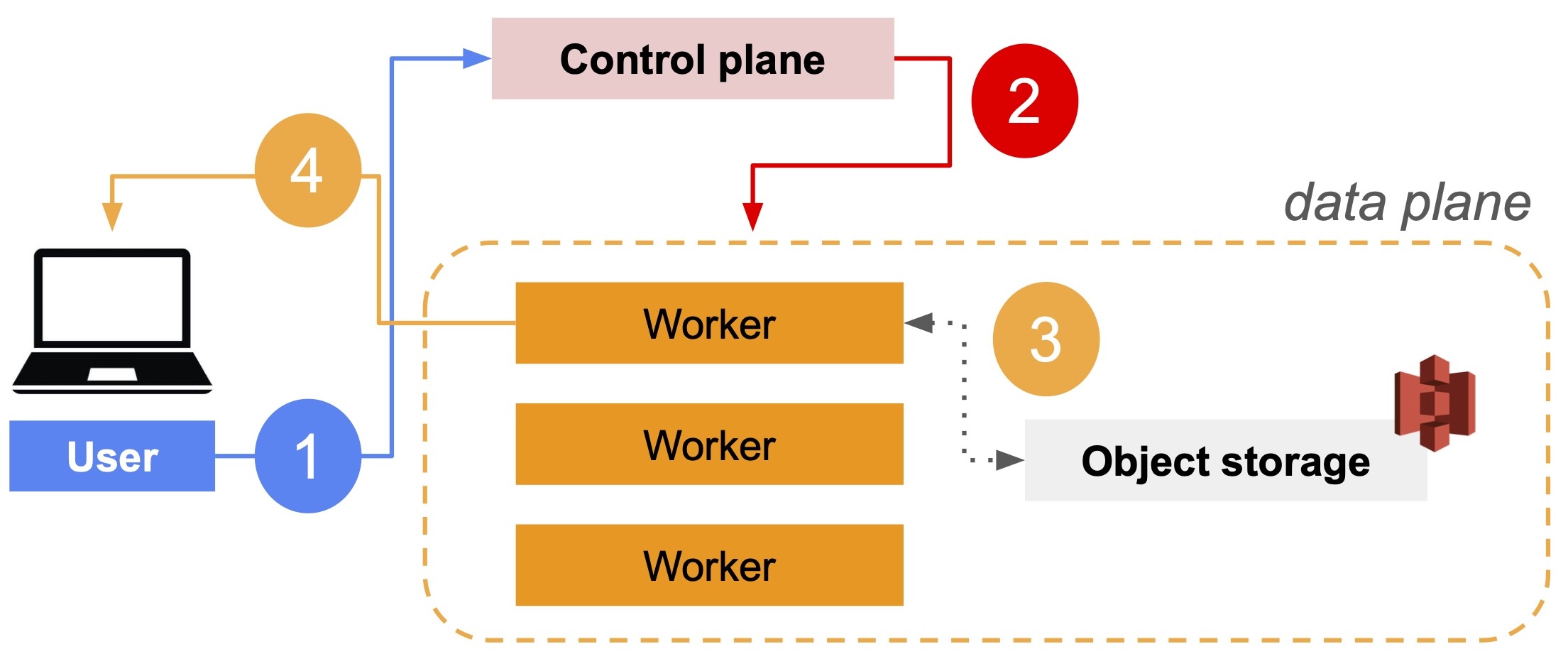}}
\caption{\textbf{A \textit{run} in Bauplan}: 1) a user (humans and / or coding assistants) writes code locally, and triggers the run; 2) the control plane parses the code into a plan and sends it to a worker for execution; 3) the worker reads/writes data from/to S3 and 4) streams logs and results to the user.}
\label{fig:cloud}
\end{figure}

Bauplan has a standard lakehouse architecture (Figure~\ref{fig:cloud}): a control plane, a data
plane, and a local client (CLI or SDK) to trigger cloud operations~\cite{10.1145/3702634.3702955}. While the architecture is straightforward, it identifies three key \emph{moments} when executing a DAG (Bauplan's \textit{run} command); ($1$) the local code environment (a user's IDE) \textit{before} a run, ($2$) the control plane (communicating DAG metadata) preparing the run (\textit{before} DAG execution begins), and finally ($3$) the worker process \textit{after} execution but before persisting data. As a general principle, we should never fail at a later moment if we could have failed at a
previous one. In other words, a ``correct-by-design'' lakehouse should not just eliminate as many failure modes as possible, but should also \emph{fail fast} when a failure occurs.

Importantly, Bauplan is layered on top of storage that already provides atomic single-table snapshot evolution and optimistic concurrency control (e.g., Apache Iceberg plus a catalog/metadata service). We do not re-prove those guarantees in this work, but assume they work as expected. Bauplan adds three higher-level correctness properties at the pipeline layer: checkable interfaces between transformations, reproducible multi-table collaboration through versioned branches and commits, and atomic publication of an entire pipeline run. In other words, Iceberg provides table-level atomicity; Bauplan provides pipeline-level structure and publication semantics. After our core design choices have been described, Section~\ref{sec:formal} will provide a deep dive on the correctness primitives.

\subsection{Programming abstractions} 
\label{sec:programming}

\textbf{Claim:} \textit{schema failures are interface bugs, so pipeline boundaries must be explicit and checkable}. 

We address schema failures by turning pipeline interfaces into explicit, machine-checkable contracts. Concretely, each node in the DAG exposes a schema object whose fields represent the columns that flow across nodes, together with their expected types. This makes the boundary between transformations explicit, as exemplified by the snippets below, which add lightweight annotations to the original DAG to elegantly achieve schema and lineage inference. For example, \texttt{col2} is propagated unchanged from \textit{Node 1} to \textit{Node 3}, while \textit{Node 2} introduces fresh columns (\texttt{col4}, \texttt{col5}), and \textit{Node 3} can legally narrow a type (e.g., \texttt{col4}: float to \texttt{col4}: int) when the transformation includes an explicit cast.

\begin{lstlisting}[
  language=Python,
  showstringspaces=false,
  columns=fullflexible,
  caption={Contracts as types. Note that $col2$ is propagated "as-is" while $col4$ is propagated from Node 2 to Node 3 but the type is \textit{narrowed}.},
  basicstyle=\ttfamily\scriptsize,
  numbers=none
]
class ParentSchema(BauplanSchema): # "Node 1"
    col1: str
    col2: datetime
    _S  : int

class ChildSchema(BauplanSchema): # "Node 2"
    col2: datetime                 # inherited type
    col4: float                    # fresh type
    col5: UNION(str, None)         # fresh type

class Grand(BauplanSchema): # "Node 3"
    col2: datetime           # inherited type
    col4: int                # inherited type is narrowed
\end{lstlisting}

\begin{lstlisting}[
  language=SQL,
  showstringspaces=false,
  columns=fullflexible,
  caption={Node \#1 as a ``typed'' transformation in a declarative language. A concise SQL
           annotation need only describe "table names", but we can explicitly associate the
           result set with a defined schema.},
  basicstyle=\ttfamily\scriptsize,
  numbers=none
]
-- parent_table: ParentSchema <- raw_table
SELECT col1, col2, SUM(col3) as _S FROM raw_table
\end{lstlisting}

\begin{lstlisting}[
  language=Python,
  showstringspaces=false,
  columns=fullflexible,
  caption={Node \#$2$ and \#$3$ as ``typed'' transformations in an imperative language (with sample business logic).},
  basicstyle=\ttfamily\scriptsize,
  numbers=none
]
# "Node 1" -> "Node 2"
def child_table(df: ParentSchema = parent_table) -> ChildSchema:
    # returns `child_table` with the schema `ChildSchema`
    return df.select([
        col('col2'),
        lit(float()).alias('col4'),
        lit(None).alias('col5'),
    ])

# "Node 2" -> "Node 3"
def grand_child(df: ChildSchema = child_table) -> Grand:
    # returns `grand_child` with the schema `Grand`
    return df.select([
        col('col2'),
        arrow_cast(col('col4'),
                   str_lit('Int64')).alias('col4'),
    ])
\end{lstlisting}

These contracts enable fail-fast behavior at the earliest possible point in the execution life-cycle. When the developer (or agent) authors code, local type checkers can catch obvious mismatches immediately. Next, before scheduling any distributed execution, the control plane can parse the DAG metadata and validate that adjacent nodes compose (e.g., every referenced column exists with a compatible type, and -- if the transformation language allows inspection -- casts are present when necessary). Finally, at the worker, runtime checks validate that the physical data actually conforms to its declared schema before any results are persisted, ensuring that late-discovered schema problems do not leak inconsistent state into storage. 

Types also give Bauplan a principled handle on data quality checks without additional tools, in line with the ``everything as code'' design principles of modern Pythonic frameworks \cite{Tagliabue2023ReasonableSM} and best practices for AI-assisted coding. For example, \texttt{col5} is explicitly declared as nullable, while \texttt{col4} is not; the runtime can therefore treat unexpected nulls as contract violations and reject outputs early. We refer the reader to the appendix for examples of column-level annotations.

\subsection{Collaboration workflows}
\textbf{Claim:} \textit{we can reuse Git's mental model for data, if the atomic versioned objects are table snapshots}. 

\textit{Git}-based workflows are considered best practices in software engineering, both in development and debugging. As traditional software is often stateless (conditional on small inputs, e.g. a JSON input), the code base is solely responsible for its behavior: new capabilities are developed and tested in isolation on \textit{feature branches}, then shipped to production through CI/CD and the usual PR flow. \textit{Conversely}, we can go in the other direction -- from production to testing -- when reproducing an issue from a deployed system in a local, controlled setup. As remarked in Section~\ref{sec:failures}, data pipelines are considerably more stateful: even if (a big \textit{if} \cite{10.1145/3702634.3702955}) we keep cloud infrastructure constant, the same code may have different behavior depending on the state of the input data in the lake: is it possible to re-use \textit{Git} abstractions to achieve reproducibility without asking humans and AIs to learn new mental models?

Our design is motivated by the observation that open table formats (i.e. Apache Iceberg~\cite{iceberg}) already provide a strong primitive: ACID-compliant, \textit{single} table snapshots, with optimistic locks guaranteed by a relational database \cite{nessie}. Since we already have a database, we can easily map every lakehouse \textit{write} recorded there (a table creation, new rows inserted by a transformation etc.) to a \textit{commit} -- an immutable, unique reference to the state of all table snapshots at that moment. From there, a Git-like ordering on commits can be induced: a \textit{branch} is then a movable reference to the \textit{HEAD} of a sequence of commits, a \textit{tag} is an immutable semantic tag referencing a commit, a \textit{merge} of a branch into another applies atomically (pending conflicts) changes from the source to the destination. If desired, a richer set of logic and conditions can be then defined: for example, fast-forward \textit{vs} three-way merges can be defined on top of table snapshots, as well as primitives such as \textit{rebase} which comes from software best practices (see the model in Section~\ref{sec:formal}). Production is then identified with the \texttt{main} branch. 

By mapping Iceberg snapshot evolution to \textit{Git} primitives, we get Git-like workflows with no additional infrastructure: as an example, the snippet below showcases the experimentation and debugging workflows backed by the underlying graph of data evolution. Note that each \textit{run} is identified uniquely with a \texttt{run\_id}, and it is associated with the state of the lake (the data commit) and the pipeline code (the local IDE folder sent to the control plane) at the start, ensuring that we can run the same code on the same input data without having to manually move between systems or keep a separate bookkeeping system \cite{10.1145/3650203.3663335}. At the physical level, implementation is simplified by the existing constraints of Iceberg tables: parquet files containing the actual data and snapshot files are immutably stored in object storage, simplifying bookkeeping and copy-on-write semantics. When a new branch is created, nothing changes in the underlying lake: only when data is added to the newly created branch, new parquet files are written and logically isolated (not unlike what happens to code files in \textit{Git}). Similarly, merge operations are only logical changes, linking physical parquet files (atomically) to a new branch, without data duplication.

\begin{lstlisting}[
  language=Python,
  showstringspaces=false,
  columns=fullflexible,
  caption={Exposing branching and debugging workflows to humans and agents as simple, Git-like APIs.},
  basicstyle=\ttfamily\scriptsize,
  numbers=none
]
import bauplan

client = bauplan.Client()
# create a feature branch from production data
feature_branch = client.create_branch('feature', from='main')
# run DAG from a local folder and get back an immutable object
run_state = client.run('DAG_code_folder/', ref=feature_branch)
print(run_state.run_id, run_state.ref, run_state.code_zip)
# experiment -> production: once the branch is reviewed, merge
assert client.merge(feature_branch, into='main')

# later, reproduce an issue from a production run_id
prod_state = client.get_run(run_id)
# start a debug branch from the run starting commit
debug_branch = client.create_branch('repro', from=prod_state.ref)
# fix the DAG, then re-run (merge if happy etc.)
run_state = client.run('repro_code_folder/', ref=debug_branch)

\end{lstlisting}

Data-level collaboration primitives cannot be overstated especially in an agentic setting. Recent evidence from real engineering workflows \cite{openai2025_scaling_code_verification} highlights that scalable adoption hinges on a resource shift from writing to reviewing and verifying changes. Bauplan's branching and time-travel abstractions help operationalize this natural division of labor: agents propose pipelines on isolated branches, while humans validate contracts and outcomes.

\begin{figure}
\centerline{\includegraphics[width=0.45\textwidth]{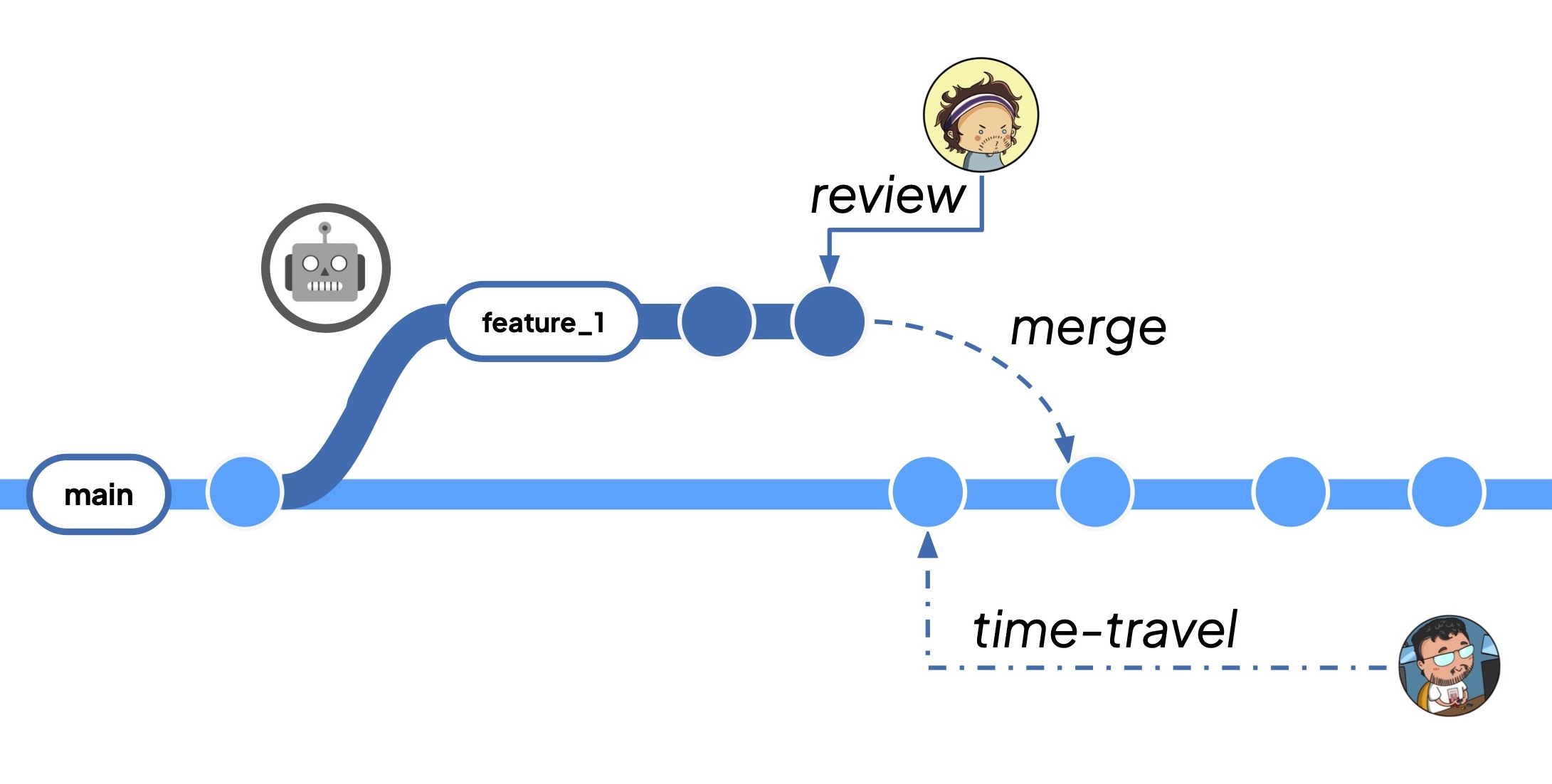}}
\caption{\textbf{Git-for-data and collaboration.} Git-like abstractions -- such as commits, branches and merges -- enable PR-based reviews and reproducibility. As argued in Section~\ref{sec:trans}, these concepts are however not sufficient for global correctness.}
\label{fig:collab}
\end{figure}

\subsection{Transactional pipelines} 
\label{sec:trans}
\textbf{Claim:} \textit{table-level atomicity is insufficient for pipeline correctness; lakehouse runs must provide \emph{pipeline-level} atomicity, i.e., publish either all outputs of a run or none}.

\begin{figure}
\centerline{\includegraphics[width=0.45\textwidth]{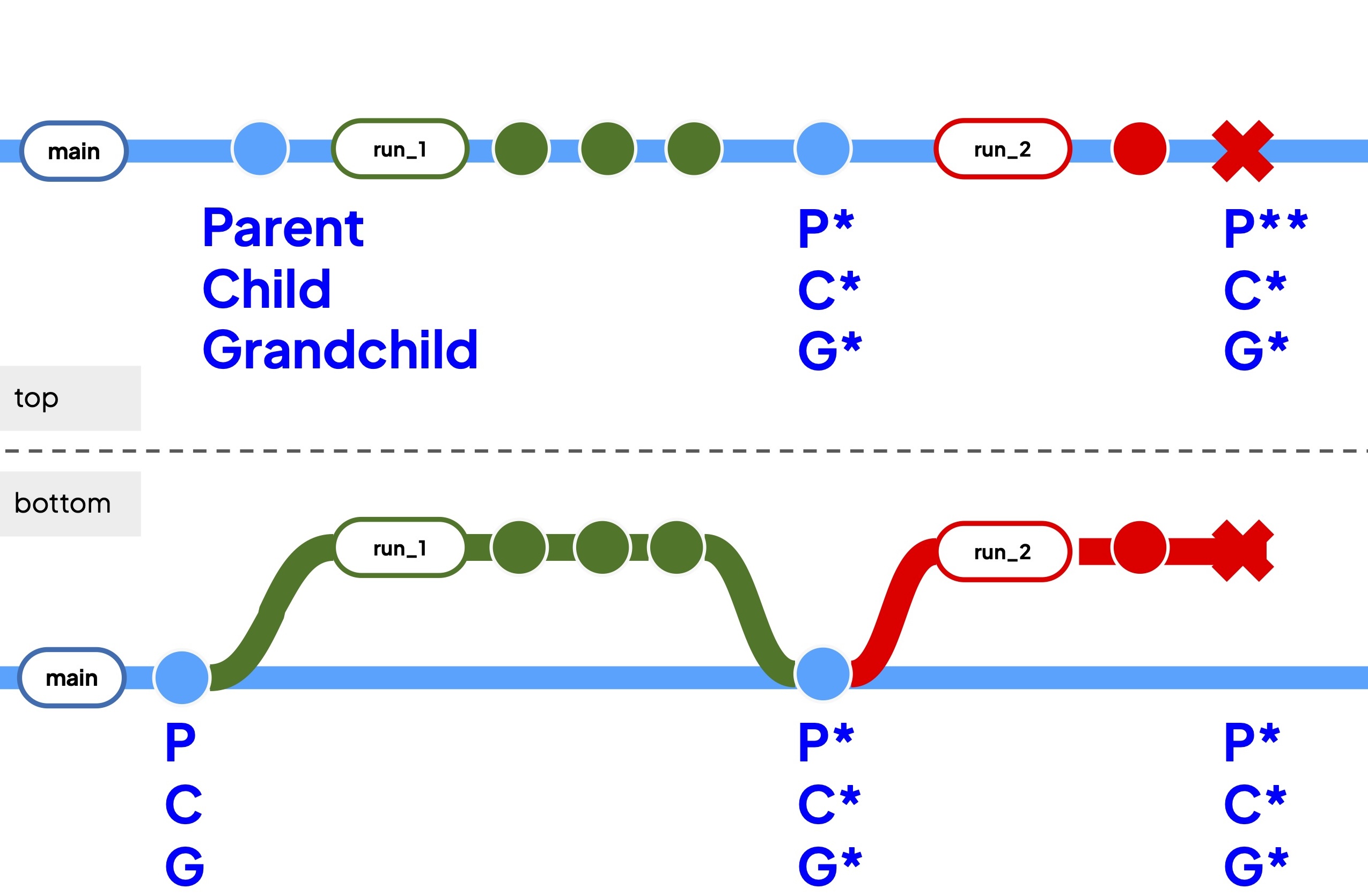}}
\caption{\textbf{Transactional pipelines}. \textit{Top}: without coupling temporary branches with runs, \texttt{run\_2} leaves \textit{main} with a new version of \(Parent\) but an old version of \(Child\) and \(Grandchild\). \textit{Bottom}: the \textit{run} API guarantees atomic publication of all tables on success, and isolation in case of failure.}
\label{fig:transaction}
\end{figure}

No primitive we have seen so far prevents the ``half-written pipeline'' from Section~\ref{sec:failures}. Uniquely identifying runs with a snapshot of input data and DAG code may solve reproducibility, but not correctness, as exemplified by Figure~\ref{fig:transaction} (top). \texttt{run\_1} executes successfully on the \texttt{main} branch, producing a commit at
each \textit{write}. Tables $P$(arent), $C$(hild) and $G$(randchild) are therefore updated to (respectively) snapshot $P*$, $C*$, $G*$; however, \texttt{run\_2} breaks after updating $P*$ to $P**$ but before updating $C*$. From the Iceberg point of view, the system is consistent, as both $P**$ and $C*$ are valid table states; however, there is a clear sense in which the system is globally \textit{inconsistent}, as the intention of whoever triggered \texttt{run\_2} was \textit{not} to get \texttt{main} into a partially stale state $\{P**, C*, G*\}$. Since
\texttt{main} can be accessed at any point by decoupled downstream systems, the global inconsistency may percolate uncontrollably without warnings. How can a dashboard know that it is reading from an incorrect state?

Creating MVCC-style \cite{10.5555/12518} transaction boundaries in the lakehouse is not straightforward. Unlike vertically integrated databases, lakehouse pipelines run on ephemeral compute and decoupled storage. Bauplan DAGs are restricted to use functions with signature \textit{Table(s) -> Table} (in line with other data engineering frameworks \cite{dbt}), but within those functions, anything can happen: any SQL query or Python version and package are allowed, as long as the signature is respected. At the programming level, we still want to pursue our general principle: we don't want actors to learn new complex abstractions, nor do we want to delegate correctness to the application layer. Our key insight is to slightly modify the semantics of the \texttt{run} API and \textit{logically} couple function execution with data branches, as exemplified by the following protocol. If $B$ is the target branch for execution, a new \texttt{run}:

\begin{enumerate}
  \item automatically creates a new ``transactional'' branch $B'$ from $B$;
  \item writes the DAG tables into $B'$ (each table commit is atomic as per Iceberg guarantees above);
  \item runs data tests / user-defined verifiers etc. on $B'$;
  \item only if no code or data error is raised, merges $B'$ back to $B$ and deletes it.
\end{enumerate}

Figure~\ref{fig:transaction} (bottom) illustrates first the happy path, with the transactional branch associated to \texttt{run\_1} atomically merged into the target branch at the end. Then, \texttt{run\_2} transactional branch illustrates the unhappy path: failure to update $C*$ to $C**$ does \textit{not} compromise
\texttt{main}, which continues to serve downstream consumers a globally consistent state of the first successful run, instead of seeing the state of a partially failed run. This is the equivalent of upgrading a ``partial failure'' into a ``total failure,'' but for data pipelines and storage instead of distributed systems. Then, as an additional bonus, the transactional branch which could not be merged (was ``aborted'')  is reachable by any user for debugging and inspection. This enables a unique capability to triage the failure of \texttt{run\_2} by investigating data quality checks and querying the faulty intermediate asset.

By slightly modifying the \texttt{run} API and re-using Git-like concepts \textit{we already adopted}, we get the best of both worlds: a transactional API that is easy to reason about and moves correctness guarantees out of users' hands, and the elasticity and scalability of a fully distributed system. While the API acts as a \textit{logical} monolith that couples execution and storage, the \textit{physical} system is elastic and scalable.

The transactional branch protocol introduces metadata and coordination overhead relative to direct writes to the target branch. In our target OLAP setting, this trade-off is acceptable because pipelines are coarse-grained, multi-table jobs whose dominant costs are typically storage I/O and compute, not branch creation or metadata updates. The protocol may be less attractive for extremely high-frequency or fine-grained update workloads; this is a deliberate design choice, and further distinguishes our target from OLTP branching systems.

We introduced and motivated our Git-like abstractions as a natural mental model that helps with both collaboration and correctness. However, the full, formal correctness of the \texttt{run} protocol above hinges on a much stricter isomorphism between data branches and multi-table transactions. Inspired by the success of lightweight formal methods~\cite{Bornholt2021} to preemptively catch bugs and unlikely race conditions in distributed systems, we turn to Alloy to push the analogy to its limits: when transactions and branches diverge (if at all)?

\section{A lightweight formal model}
\label{sec:formal}
To stress-test the intuition that transactions are essentially branches, we formalize a \emph{Git-for-data} core in Alloy \cite{DBLP:books/daglib/0024034}. The model is presented as a minimal lakehouse formalization, focused on the interplay between commits, branches and runs, and abstracting away the underlying ACID considerations at the Iceberg level. In other words, \textit{assuming} the existence of single-table snapshot evolution, the model tests whether pipeline transactions can be represented by Git primitives.

We explain the core modelling choices, before highlighting interesting counterexamples we found by running the model (see Appendix~\ref{appendix:code} for the code).

\paragraph{Commits and branches.}
A commit contains a mapping from tables to snapshots and a parent relation; a branch points to exactly one commit. The system starts with a single branch \texttt{Main} and a single root commit (\texttt{Init}).

\begin{lstlisting}[
  caption={Commits as lake states and branches as movable pointers.},
  label={lst:alloy-core-min},
  basicstyle=\ttfamily\scriptsize,
  numbers=none
]
sig Table {}
sig Snapshot {}

var sig Commit {
    var tables: Table -> lone Snapshot,
    var parent: set Commit
}

var sig Branch {
    var commit: one Commit
}

var one sig Main extends Branch {} { this' = this }
\end{lstlisting}

\paragraph{The only state-changing write.} Without loss of generality, we represent all transformations as opaque, atomic operations on target tables.
All lake evolution is funneled through \texttt{createTable[b,t]}, which (when \texttt{t} does not yet exist on \texttt{b}) allocates a fresh snapshot and a fresh commit \texttt{co}, sets \texttt{co.parent} to the previous head, and advances \texttt{b.commit} to \texttt{co}.

\begin{lstlisting}[
  caption={The only mutating operation: create a new snapshot and a new commit, then advance the branch head.},
  label={lst:alloy-createTable},
  basicstyle=\ttfamily\scriptsize,
  numbers=none
]
pred createTable[b: Branch, t: Table] {
    no b.commit.tables[t]

    one s: Snapshot, co: Commit' - Commit {
        no tables.s
        Commit' = Commit + co
        tables' = tables + (co -> (b.commit.tables + t -> s))
        parent' = parent + (co -> b.commit)
        commit' = commit ++ (b -> co)
    }
    Branch' = Branch
}
\end{lstlisting}

\paragraph{Runs as step-by-step writes on a chosen branch.}
A pipeline is a sequence of tables (\texttt{plan: seq Table}). Each successful step applies \texttt{createTable} to the next planned table and advances an internal counter, \texttt{idx}.

\begin{lstlisting}[
  caption={Runs: begin on a branch, execute \texttt{createTable} step-by-step, then finish or fail (excerpt).},
  label={lst:alloy-run-min},
  basicstyle=\ttfamily\scriptsize,
  numbers=none
]
sig Pipeline { plan: seq Table }

sig Run {
    pipeline   : one Pipeline,
    var lastCommit : lone Commit
}

pred begin[r: Run, b: Branch] { ... }
\end{lstlisting}

\paragraph{Minimal counterexamples.}

\begin{figure}
\centerline{\includegraphics[width=0.45\textwidth]{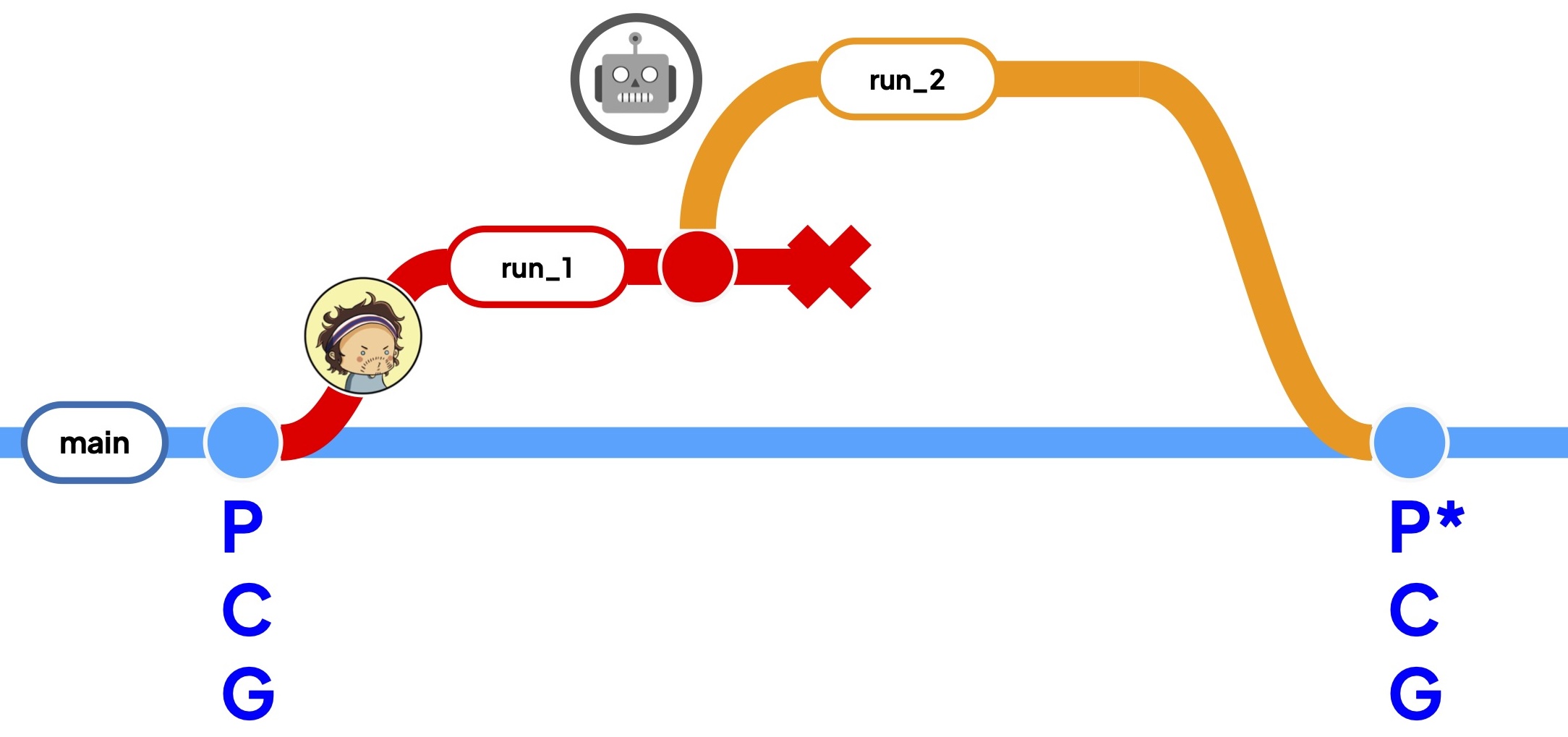}}
\caption{\textbf{A counterexample.} A user's run failed after the first commit, leaving an aborted branch open. An agent can do arbitrary work by branching off that commit: if a merge back to \texttt{main} happens from \texttt{run\_2}, the system ends up in an inconsistent state with respect to \texttt{run\_1} semantics.}
\label{fig:counter}
\end{figure}

We establish minimal adequacy for the model by first providing commands to reproduce the asymmetry between \textit{top} and \textit{bottom} in Figure~\ref{fig:transaction}. We then leverage Alloy's model checking capabilities to produce counterexamples that highlight missing nuances in our concepts. Figure~\ref{fig:counter} depicts a degenerate system evolution that is compatible with the above concepts and constraints, but it is still producing an inconsistent state. A failed, aborted run by a user results in a ``dangling'' branch that is not merged back -- so far so good. However, in a distributed system such as Bauplan, other actors may see that branch as available, and decide to work off it: when that second branch is merged onto the original branch, a global inconsistency is back -- the original user never meant for that state to be available to downstream systems. 

We could rephrase the key conceptual insight from our formal model as such: arbitrary branches -- but not necessarily arbitrary transactions -- are nestable, and actors can create new branch from existing ones at any point in time if no additional guardrails are provided. It is important to note that simply making aborted branches invisible after failure (as it happens in a transactional database such as \textit{Postgres}) may solve this counterexample, at the price of precluding more complex \textit{legal} workflows that Bauplan could plan to optimize resource usage. If \texttt{child} logic was wrong but the DAG is deemed to be idempotent, Bauplan could plan a re-run with new \texttt{child} code by starting from the already materialized \texttt{parent}, instead of re-calculating it -- in other words, under certain conditions, an aborted transactional branch could be used as a starting branch for non-aborted runs. We plan on refining our model further, leveraging these and more insights from lightweight formal methods to inform API-level changes and trade-off expressivity and guarantees in a principled manner.

\section{Related work}

\paragraph{DAG frameworks and abstractions.} 
Unlike Bauplan, popular DAG frameworks are more limited in expressivity and not vertically integrated into the data stack, resulting in limited guarantees. \textit{Dagster}~\cite{web:dagster} (mostly Python-centric) and \textit{dbt}~\cite{web:dbt} (mostly SQL-centric) offer no transactional APIs but have schema checks on assets: however, checks only work at runtime \textit{and} on assets that have been persisted, with no attempt to lift contract composition into a standardized type checking. Prefect~\cite{web:prefect} allows users to define a $transaction$ context manager that provides failure and rollback handlers. Prefect does not control the underlying data layer, which leaves users to manually define appropriate atomic data operations within the context manager, with the associated and well-documented risks \cite{10.1145/3638553}.

\paragraph{MVCC and Git modelling.}
Our work aligns with recent database literature highlighting similarities between Git primitives and transactions. While TARDIS~\cite{10.1145/2882903.2882951} proposes branch-and-merge for weak isolation in a key-value store, Bauplan's APIs apply to distributed storage in a multilanguage OLAP setting, re-using collaboration abstractions for transaction boundaries. Branching systems for OLTP databases \cite{dolt} are related but address a different problem, as they  operate over transactional row-store workloads.

While previous work suggests that Git commit graph is a form of transaction \cite{yilmazdittrich}, our contribution targets a different correctness boundary: data pipelines are multi-table and multi-language, and the dominant failure mode is \textit{pipeline-level} partial publication rather than tuple-level anomalies. Moreover, our Alloy counterexample highlights that the additional expressivity enabled by nested branching can be a safety hazard for pipeline transactions unless transactional branches are given stricter visibility. As part of future work, we plan on modelling isolation levels in Bauplan following the formalization in \cite{cerone_et_al:LIPIcs.CONCUR.2015.58}.

\section{Conclusion and future work}

We described Bauplan, a code-first lakehouse, and our work-in-progress towards a formal model of its correctness. As of today, humans and coding assistants have created millions of data branches on it, making Bauplan the first system of this kind tested at industry scale. As our north star, we aim for a \textit{correct-by-design} lakehouse, one in which failures become unrepresentable: ill-typed pipelines should not be planned, inconsistent plans should not be run, failed runs should not be published. As we plan to refine our type inference and Alloy model, we are reminded of the words of a wise man at the dawn of AI: ``we can only see a short distance ahead, but we can see plenty there that needs to be done''.


\bibliographystyle{ACM-Reference-Format}
\bibliography{base}

\appendix

\section{Extended Contracts}
\paragraph{Binary DAG Node with Lineage} On top of table-level data contracts and dependencies, \textit{Bauplan} supports annotations for column inheritance. This can be used to fail a data pipeline early, when a column's type is altered unexpectedly; or, this can be used to analyze properties of a column's usage across a DAG, identifying when the column's type is changed or providing insight about how the column is used.

In this snippet, we show how $FriendSchema$ explicitly declares that it inherits columns $col2$  and $col5$ from $ChildSchema$ and $col4$ from $Grand$. This schema inherits columns from multiple inputs, which is represented by the parameters of the $family_friend$ DAG node. Finally, $FriendSchema$ is explicit about its expectation to alter the type metadata of $col5$ to be ``not null'', meaning all $None$ values will be filtered out.

\begin{lstlisting}[
  language=Python,
  showstringspaces=false,
  columns=fullflexible,
  caption={An output schema that illustrates explicit column inheritance from two input schemas.},
  basicstyle=\ttfamily\scriptsize,
  numbers=none
]
# Schema for family friend
class FriendSchema(BauplanSchema): # "Node 4"
    col2 = ChildSchema.col2          # inherited type
    col4 = Grand.col4                # inherited type
    
    # inherited type; explicitly not null
    col5 = ChildSchema.col5[NotNull]
\end{lstlisting}

\begin{lstlisting}[
  language=Python,
  showstringspaces=false,
  columns=fullflexible,
  caption={A DAG Node that illustrates a binary transformation (two inputs and one output)},
  basicstyle=\ttfamily\scriptsize,
  numbers=none
]
# "Node 2", "Node 3" -> "Node 4"
def family_friend(
    df_child: ChildSchema = child_table,
    df_grand: Grand       = grand_child,
) -> FriendSchema:
    df_grand = df_grand.select(
      col('col2'),
      col('col4').alias('4_grand'),
    )

    # returns table with schema `FriendSchema`
    return (
            df_child
            .join(df_grand, on=["col2"], how="inner")
            .filter(
                col("col5").is_not_null()
                & ((col("4_grand") - col("col4")) < 0.5)
            )
            .select(
                col("col2"),
                col("4_grand").alias("col4"),
                col("col5"),
            )
        )
\end{lstlisting}

Column properties (such as ``nullability'') must be validated by a worker process where the data is available, but that validation can occasionally be optimized by the control plane. Naively, after the worker process validates that the result of node $family_friend$ has no null values in $col5$, then additional validation after subsequent DAG nodes can be omitted in the following cases: ($1$) output schemas of subsequent DAG nodes are defined and trusted, ($2$) subsequent DAG nodes use a transformation language that can be inspected, such as SQL, and the transformation is guaranteed to maintain the column's nullability. In other words, we could be able to statically prove \textit{pre} and \textit{post} conditions (``Dafny-style'') for certain DAG nodes, i.e. proving without running any code that if some data quality traits are there in the input, they must necessarily be there in the output as well.

\section{Open-source code}
\label{appendix:code}

Open-source Alloy code to reproduce our findings is available at: \url{https://github.com/BauplanLabs/git_for_data}.

\end{document}